\documentclass[twocolumn]{27onr_art}
\usepackage{ulem}
\usepackage{harvard}
\usepackage{graphicx}

\pdfoutput = 1

\def\be{\begin{eqnarray}}
\def\ee{\end{eqnarray}}
\def\benl{\begin{eqnarray*}}
\def\eenl{\end{eqnarray*}}

\newcommand{\nwc}{\newcommand}
\nwc{\bm}{\boldmath}
\nwc{\m}{\mbox}
\nwc{\ubm}{\unboldmath}
\nwc{\bmU}{\m{\bm$U$\ubm}}
\nwc{\bmX}{\m{\bm$X$\ubm}}
\nwc{\bmu}{\m{\bm$u$\ubm}}
\nwc{\bmx}{\m{\bm$x$\ubm}}
\nwc{\bmz}{\m{\bm$z$\ubm}}
\nwc{\bmv}{\m{\bm$v$\ubm}}
\nwc{\bmw}{\m{\bm$w$\ubm}}
\nwc{\bmW}{\m{\bm$W$\ubm}}
\nwc{\bmn}{\m{\bm$n$\ubm}}
\nwc{\bmG}{\m{\bm$G$\ubm}}
\nwc{\bmF}{\m{\bm$F$\ubm}}
\nwc{\bmI}{\m{\bm$I$\ubm}}
\nwc{\bmN}{\m{\bm$N$\ubm}}
\nwc{\bmP}{\m{\bm$P$\ubm}}
\nwc{\bmcalP}{\m{\bm $\cal P$\ubm}}
\nwc{\bmV}{\m{\bm$V$\ubm}}
\nwc{\bmS}{\m{\bm$S$\ubm}}

\begin{document}

\title{A Numerical Formulation for Simulating Free-Surface Hydrodynamics}

\author{Thomas T. O'Shea, Kyle A. Brucker, \\ Douglas G.\ Dommermuth, and Donald C. Wyatt}

\affiliation{\small Naval Hydrodynamics Division, Science Applications International Corporation,
\\ 10260 Campus Point Drive, MS C5, San Diego, CA  92121}

\maketitle

\begin{abstract}
Cartesian-grid methods in combination with immersed-body and volume-of-fluid methods are ideally suited for simulating breaking waves around ships.  A surface panelization of the ship hull is used as input to impose body-boundary conditions on a three-dimensional cartesian grid.  The volume-of-fluid portion of the numerical algorithm is used to capture the free-surface interface, including the breaking of waves.  The numerical scheme is implemented on a parallel computer.  Recent improvements to the numerical scheme are discussed, including implementation of a new multigrid procedure and conversion to MPI communication.   Numerical predictions are compared to laboratory measurements of a towed transom-stern model.
\end{abstract}

\section{Introduction}

\citeasnoun{fureport08} have performed towing-tank experiments for a canonical transom-stern model for four different tow speeds.  The objectives of the experiments are to improve understanding of transom-stern flows, provide validation of CFD methods, provide validation of a new transom-stern theory, and provide input to bubble modeling.  The laboratory measurements include free-surface elevations using surface imaging techniques and wave-cut methods, void fraction, bubble size and velocities, flow velocities, and turbulence including the formation of coherent structures.   Preliminary laboratory results are compared to numerical predictions using the Numerical Flow Analysis (NFA) code.   There is good agreement between measurements and predictions for drag force and wave cuts.

The NFA code provides turnkey capabilities to model breaking waves around a ship, including both plunging and spilling breaking waves, the formation of spray, and the entrainment of air.   NFA uses a cartesian-grid formulation with immersed-body and volume-of-fluid (VOF) methods.  The governing equations are formulated on a cartesian grid thereby eliminating complications associated with body-fitted grids. The sole geometric input into NFA is a surface panelization of the ship hull.  No additional gridding beyond what is already used in potential-flow methods and hydrostatics calculations is required.   The ease of input in combination with a flow solver that is implemented using parallel-computing methods permit the rapid turn around of numerical simulations of complex interactions between free surfaces and ships.   Details of the numerical formulation are provided in \citeasnoun{dommermuth06} and \citeasnoun{dommermuth07}.

Recent enhancements to NFA include a new multigrid solver that improves the rate of convergence of the flow solver and new MPI communication algorithms that improve communication between processors on a parallel computer.   These two improvements are a step toward simulating very large CFD problems with order one billion grid points.   This grid resolution is required to span the wide range in scales in both space and time that characterize the current generation of naval ships.   

The results of  several other NFA simulations are reported in the proceedings of this symposium.   \citeasnoun{fu08} perform preliminary studies of wave impact.   NFA predictions of a plunging breaking wave agree very well with measurements for free-surface elevations and water-particle velocities.     \citeasnoun{ratcliffe08} continue their earlier studies of incident waves interacting with towed model.    Numerical predictions of the diffracted waves agree with QViz measurements in the bow region for a high-speed case with a large incident wave.   \citeasnoun{wyatt08} compare full-scale lidar measurements of the rooster-tail region behind the R/V Athena to numerical predictions using NFA.     Mean free-surface elevations are in good agreement, but free-surface spectrums differ at higher frequencies.

\section{\label{sec:formulation}Formulation}

Consider turbulent flow at the interface between air and water.  Let $u_i$ denote the three-dimensional velocity field as a function of space ($x_i$) and time ($t$).  The coordinate system is fixed.    For an incompressible flow, the conservation of mass gives
\begin{eqnarray}
\label{mass}
\frac{\partial u_i}{\partial x_i} = 0 \;\; .
\end{eqnarray}
\noindent  $u_i$ and $x_i$ are normalized by $U_o$ and $L_o$, which denote the free-stream velocity and the length of the body,
respectively.

Following a procedure that is similar to \citeasnoun{rider94}, we let $\phi$ denote the fraction of fluid that is inside a cell. By
definition, $\phi=0$ for a cell that is totally filled with air, and $\phi=1$ for a cell that is totally filled with water.

The advection of $\phi$ is expressed as follows:
\begin{eqnarray}
\label{vof}
\frac{\partial \phi}{\partial t}+ \frac{\partial}{\partial x_j} \left(u_j \phi \right)= \frac{\partial Q}{\partial x_j} \;\; ,
\end{eqnarray}
\noindent  $Q$ is a sub-grid-scale flux that can model the entrainment of gas into the liquid.  \citeasnoun{dommermuth98} provide an example of a sub-grid model.   Since the present formulation maintains a sharp interface, $Q=0$.

Let $\rho_\ell$ and $\mu_\ell$ respectively denote the density and dynamic viscosity of water. Similarly, $\rho_g$ and $\mu_g$ are the corresponding properties of air.  The flows in the water and the air are governed by the Navier-Stokes equations:
\begin{eqnarray}
\label{navi}
\frac{d u_i}{d t}+\frac{\partial}{\partial x_j} \left(u_j u_i \right)  =  -\frac{1}{\rho} \frac{\partial P}{\partial x_i}  \nonumber  \\
+\frac{1}{\rho R_e} \frac{\partial}{\partial x_j} \left( 2 \mu S_{ij} \right) -\frac{\delta_{i3}}{F_r^2}  +\frac{\partial  \tau_{ij}}{\partial x_j} \;\; ,
\end{eqnarray}
\noindent where $R_e=\rho_\ell U_o L_o/\mu_\ell$ is the Reynolds number and $F_r^2 = U_o^2/(g L_o)$ is the Froude number. $g$ is the acceleration of gravity.  $P$ is the pressure.  $\delta_{ij}$ is the Kronecker delta function.   $\tau_{ij}$ is the subgrid-scale stress (SGS) tensor.   An example of a SGS closure for the convective terms is provided by \citeasnoun{dommermuth98}.   $S_{ij}$ is the deformation tensor:
\begin{eqnarray}
S_{ij} & = & \frac{1}{2} \left( \frac{\partial u_i}{\partial x_j} +\frac{\partial u_j}{\partial x_i} \right) \;\; .
\end{eqnarray}
\noindent $\rho$ and $\mu$ are respectively the dimensionless variable densities and viscosities:
\begin{eqnarray}
\label{density}
\rho(\phi) & = & \lambda + (1 - \lambda) {\rm H} (\phi) \nonumber \\ \mu(\phi) &
= & \eta + (1 - \eta ) {\rm H} (\phi) \;\; ,
\end{eqnarray}
\noindent where $\lambda = \rho_g/\rho_\ell$ and $\eta = \mu_g/\mu_\ell$ are the density and viscosity ratios between air and water.
For a sharp interface, with no mixing of air and water, ${\rm H}$ is a step function.  In practice, a mollified step function is used to
provide a smooth transition between air and water.

A no-flux condition is imposed on the surface of the ship hull:
\begin{eqnarray}
\label{eqn:neumann}
u_i n_i = v_i n_i
\end{eqnarray}
\noindent $v_i$ is the velocity of the ship.  $v_i$ includes the effects of rigid-body translation and rigid-body rotation.  $n_i$ denotes the normal to the ship hull that points into the fluid.

As discussed in \citeasnoun{dommermuth98}, the divergence of the momentum equations (\ref{navi}) in combination with the
conservation of mass (\ref{mass}) provides a Poisson equation for the dynamic pressure:
\begin{eqnarray}
\label{pois} 
\frac{\partial}{\partial x_i} \frac{1}{\rho} \frac{\partial
P}{\partial x_i} = \Sigma \;\; ,
\end{eqnarray}
\noindent where $\Sigma$ is a source term.  As shown in the next section, the pressure is used to project the velocity onto a
solenoidal field.   If $Q\neq0$ in equation (\ref{vof}), then $\Sigma$ includes a sink-like term that accounts for mass diffusion.

\subsection{Numerical Time Integration}

Based on \citeasnoun{sussman03a}, a second-order Runge-Kutta scheme is used to integrate with respect to time the field equations
for the velocity field.  Here, we illustrate how a volume of fluid formulation is used to advance the volume-fraction function.  Similar examples are provided by \citeasnoun{rider94}.  During the first stage of the Runge-Kutta algorithm, a Poisson equation for the pressure
is solved:
\begin{eqnarray}
\label{eqn:poisson1}
\frac{\partial}{\partial x_i} \frac{1}{\rho(\phi^k)} \frac{\partial P^*}{\partial x_i} =\frac{\partial}{\partial x_i} \left(
\frac{u^k_i}{\Delta t}+R_i \right) \;\; ,
\end{eqnarray}
where $R_i$ denotes the nonlinear convective, hydrostatic, viscous, sub-grid-scale, and body-force terms in the momentum equations.
$u^k_i$ and $\rho^k$ are respectively the velocity components at time step $k$.  $\Delta t$ is the time step.  $P^*$ is the
first prediction for the pressure field.

For the next step, this pressure is used to project the velocity onto a solenoidal field. The first prediction for the
velocity field ($u^*_i$) is
\begin{eqnarray}
\label{eqn:runge1}
u^*_i=u^k_i+\Delta t \left( R_i-\frac{1}{\rho(\phi^k)}\frac{\partial P^*}{\partial x_i} \right)
\end{eqnarray}
The volume fraction is advanced using a volume of fluid operator (VOF):
\begin{eqnarray}
\phi^*=\phi^{k}- {\rm VOF} \left( u^k_i,\phi^k,\Delta t \right)
\end{eqnarray}
A Poisson equation for the pressure is solved again during the second stage of the Runge-Kutta algorithm:
\begin{eqnarray}
\label{eqn:poisson2}
\frac{\partial}{\partial x_i} \frac{1}{\rho(\phi^*)} \frac{\partial P^{k+1}}{\partial x_i}=\frac{\partial}{\partial x_i} \left(
\frac{u^*_i+u^k_i}{\Delta t}+R_i \right)
\end{eqnarray}
$u_i$ is advanced to the next step to complete one cycle of the Runge-Kutta algorithm:
\begin{eqnarray}
\label{eqn:runge2}
u^{k+1}_i=\frac{1}{2} \left( u^*_i + u^k_i +\Delta t \left( R_i -\frac{1}{\rho(\phi^*)}\frac{\partial P^{k+1}}{\partial x_i}
\right) \right) \;\; ,
\end{eqnarray}
and the volume fraction is advanced to complete the algorithm:
\begin{eqnarray}
\phi^{k+1}=\phi^k- {\rm VOF} \left( \frac{u^*_i+u^k_i}{2},\phi^{k},\Delta t \right)
\end{eqnarray}

\subsection{Enforcement of Body Boundary Conditions}

A no-flux boundary condition is imposed on the surface of the body using a finite-volume technique.  A signed distance function $\psi$ is used to represent the body.  $\psi$ is positive outside the  body and negative inside the body. The magnitude of $\psi$ is the minimal distance between the position of $\psi$ and the surface of the body.  $\psi$ is zero on the surface of the body.   $\psi$ is calculated using a surface panelization of the hull form.   Green's theorem is used to indicate whether a point is inside or outside the body, and then the shortest distance from the point to the surface of the body is calculated.  Triangular panels are used to discretize the surface of the body.  The shortest distance to the surface of the body can occur on either a surface, edge, or vertice of a triangular panel.  Details associated with the calculation of $\psi$ are provided in \citeasnoun{sussman01}.

Cells near the ship hull may have an irregular shape, depending on how the surface of the ship hull cuts the cell.  On these irregular boundaries, the finite-volume approach is used to impose free-slip boundary conditions.   Let $S_b$ denote the portion of the cell whose surface is on the body, and let $S_o$ denote the other bounding surfaces of the cell that are not on the body.   Gauss's theorem is applied to the volume integral of  equation \ref{eqn:poisson1}:
\begin{eqnarray}
\label{eqn:integral1}
\int_{S_o+S_b} ds  \frac{n_i}{\rho(\phi^k)} \frac{\partial P^*}{\partial x_i}  = \nonumber \\ 
\int _{S_o+S_b} ds \left( \frac{u^k_i n_i}{\Delta t}+R_i n_i \right) \;\; .
\end{eqnarray}
\noindent Here, $n_i$ denotes the components of the unit normal on the surfaces that bound the cell.   Based on equation \ref{eqn:runge1}, a Neumann condition is derived for the pressure on $S_b$ as follows:  
\begin{eqnarray}
\label{eqn:bc1}
\frac{n_i}{\rho(\phi^k)}\frac{\partial P^*}{\partial x_i} =-\frac{u^*_i n_i}{\Delta t} +\frac{u^k_i n_i}{\Delta t}+R_i n_i \;\; .
\end{eqnarray}
\noindent The Neumann condition for the velocity (\ref{eqn:neumann}) is substituted into the preceding equation to complete the Neumann condition for the pressure on $S_b$:
\begin{eqnarray}
\label{eqn:bc2}
\frac{n_i}{\rho(\phi^k)}\frac{\partial P^*}{\partial x_i}  =  -\frac{v^*_i n_i}{\Delta t} +\frac{u^k_i n_i}{\Delta t}+R_i n_i \;\; .
\end{eqnarray}
\noindent This Neumann condition for the pressure is substituted into the integral formulation in equation \ref{eqn:integral1}:
\begin{eqnarray}
\label{eqn:integral2}
\int_{S_o} ds  \frac{1}{\rho(\phi^k)} \frac{\partial P^*}{\partial x_i} n_i & = & \int _{S_o} ds \left( \frac{u^k_i n_i}{\Delta t}+R_i n_i \right) \nonumber \\
& + & \int _{S_b} ds \frac{v^*_i n_i}{\Delta t}
\end{eqnarray}
This equation is solved using the method of fractional areas.  Details associated with the calculation of the area fractions are provided in \citeasnoun{sussman01} along with additional references.  Cells whose cut volume is less than 25\% of the full volume of the cell are merged with neighbors.  The merging occurs along the direction of the steepest gradient of the signed-distance function $\psi$. This improves the conditioning of the Poisson equation for the pressure.   As a result, the stability of the  projection operator for the velocity is also improved (see equations \ref{eqn:runge1} and \ref{eqn:runge2}). 

\section{Numerical efficiency}

The capability to simulate free-surface hydrodynamics requires the ability to simulate a wide range of scales in both space and time with a large number of grid points.   The numerical solver should scale linearly with the number of unknowns and the number of CPU's.   In the next two sections, we discuss upgrades to the Poisson solver and to the communication between processors that enable very large-scale applications with order one billion grid points.

\subsection{Poisson solver}

\citeasnoun{dommermuth06} use a preconditioned conjugate-gradient method to solve the Poisson equation for the pressure (\ref{eqn:integral2}).   As the number of grid cells increase, the rate of convergence of the conjugate-gradient solver tends to stall.  A new multigrid solver has been developed to address this issue.    Multigrid uses a fine grid to reduce high-wavenumber residual errors and a coarse grid to reduce low-wavenumber errors in the residual.  A key aspect of multigrid requires moving metrics and associated variables from fine grids to coarse grids during the restriction phase \cite{wesseling}.  In Equation \ref{eqn:integral2}, the density and cut-surface areas are required on the coarser grids.   They can be moved independently or their product can be moved as one unit.   The density is located at cell centers and the cut-surface areas are located at cell faces.   If the density is moved from cell centers on the fine grid to cell centers on the coarse grid, the sharp interface between air and water is excessively smoothed.   As a result, the rate of convergence of the multigrid algorithm is reduced.   Less smoothing occurs, if the density is first moved to cell faces before the restriction phase.   In NFA, the product of one over the density times the cut-surface areas is evaluated on cell faces on the finest grid before moving it to the coarse grids.   This improves convergence of the multigrid algorithm for sharp interface methods.    Unlike the old preconditioned conjugate-gradient solver, the performance of the new multigrid method does not degrade as the number of grid points increases beyond ten million grid cells.

\subsection{Parallel computing}

Scaling, in this paper, refers to the ability of a parallel code to make adequate use of all of the processors on which it is being run. Ideally, when a massively parallel code is run on double the number processors, it would finish in half the amount of clock time. Realistically, there will be some amount of overhead associated with communicating data among processors and the time will not quite be halved while doubling processor count. 

High grid densities are increasingly needed to resolve the small-scale phenomena that are present in many current hydrodynamic calculations. This increase in grid resolution necessitates increasing the number of processors on which NFA is run to keep simulation time down to a reasonable level.   NFA's communication package has been recently upgraded to run effectively with this higher processor count. The original communication routines were written using the CRAY's native package called shmem (shared memory.) Shmem routines are optimized for large contiguous data transfers and thus are not ideal for passing guard or ghost cells between blocks in the domain decomposition method that is used by NFA. 

A new version of the the communication routines has been written using MPI-2 (message passing interface.)  The set of routines is optimized for non-contiguous data transfer through the creation of strided vector types.  Strided vector types provide information to the MPI implementation regarding exactly how data to be transferred is arranged in an array. This allows NFA to pass only the minimum amount of data necessary between blocks. 

The upgrade of the communication package has provided a unique opportunity to additionally overhaul the rest of the code. The entire code has been ported from Fortran 77 to Fortran 90.   All common blocks have been removed and replaced with modules and allocatable arrays. The inputs to the code have been taken out of include files and are now input by the user. As a result,  the code does not need to be recompiled for every run and can be executed using simple scripts.  

The file input and output in NFA has also been updated using MPI parallel file I/O. Originally when files had been read in and out of NFA,  the master process would have to collect or distribute all the data to each individual process and perform all file I/O itself. The new version of NFA uses MPI commands to allow each processor to read from or write to its own memory offset location in a common file. With no communication to perform, the files are written significantly faster. 

The development of this new version of NFA has taken place on a recently constructed LINUX cluster. The cluster houses 32 AMD quad core Opteron CPU's and an infiniband interconnect to handle all communication between processors. Inifiniband is optimized for remote memory access or RMA.  One-sided communication calls such as MPI\_get and MPI\_put make use of this and allow the processor whose memory is being accessed to continue working on its subsection of the domain while the inifiniband hardware handles most of the communication overhead.

\begin{figure}
\begin{center}
\includegraphics[width=\linewidth]{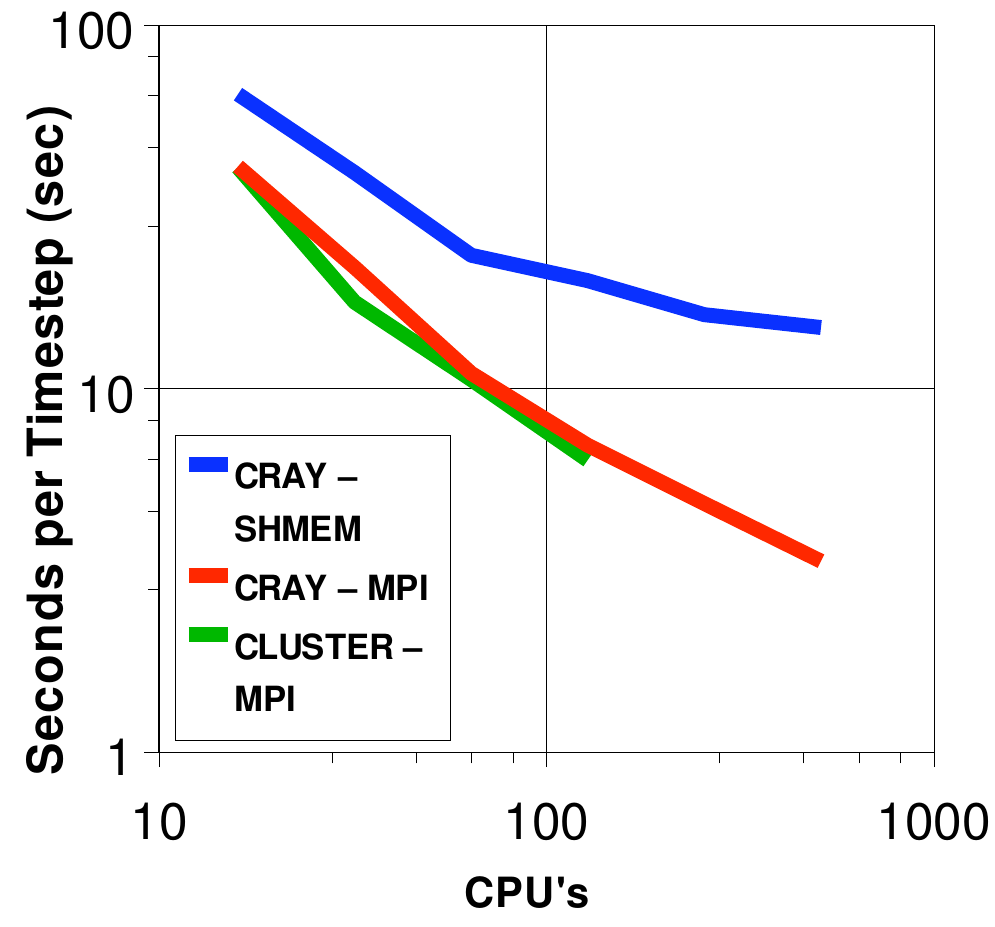}
\caption{\label{scaling} Scaling of NFA.}
\end{center}
\end{figure}

As a result of these upgrades, the scaling performance of NFA is greatly increased. Figure \ref{scaling} shows a plot of NFA's clock time per time step while solving a domain of 16.8 million cells. The data is plotted using log-log axes; consequently, a slope of negative one would indicate perfect scaling, since the doubling of CPU's would halve the computation time. There is a significant improvement between the old shmem version of NFA and the new MPI version especially as CPU count reaches 512, which shows a speed up that is almost four times the old version.   More importantly, the roll off as the number of CPU's increases is much less severe for the new MPI version relative to the old shmem version.  It should be noted that the same MPI version of NFA is running on both the CRAY and the LINUX Cluster attesting to the portability of the MPI-2 standard.

\section{Results} \label{sec:results}

NFA predictions of of the free-surface elevations near a transom-stern model moving with constant forward speed are compared to laboratory measurements.
Table \ref{table:transom} provides details of the transom-stern model tests, including the length of the model, the depth of the transom, the speed of the model, and the Froude number.  All length and velocity scales are respectively normalized by the model's length ($L_o$) and speed ($U_o$) .  The three-dimensional numerical simulations use 768x192x160= 23,592,960 grid points, 24x6x5=720 sub-domains, and 180 nodes on a Cray XT3.  The length, half width, depth, and height of the computational domain are respectively 3.0, 0.84917, 0.66667, 0.5 ship lengths ($L_o$).   These dimensions match the cross section of the DTMB towing tank.   A symmetry boundary condition is imposed on the plane $y=0$.  Grid stretching is employed in all directions.  Details of the grid-stretching algorithm are provided in \citeasnoun{dommermuth06}.  The smallest grid spacing is 0.002  near the ship and mean waterline, and the largest grid spacing is 0.007  in the far field.   The fore perpendicular and transom are respectively located at $x=0$ and $x=-1$.  The numerical simulations are slowly ramped up to full speed.   The period of adjustment is $T_o=0.5$ \cite{dommermuth06}.   Mass conservation is ensured using the regridding algorithm that is implemented by \citeasnoun{dommermuth06}.  The density is not smoothed, but a clipping algorithm is implemented \cite{dommermuth06}.  For this simulation, the non-dimensional time step is $\Delta t$=0.00025.  The numerical simulation runs 20,000 time steps corresponding to 5 ship lengths.  The simulation requires about 23 hours of wall-clock time.

\begin{table}
\begin{center}
\begin{tabular}{|l|l|l|l|} \hline
Length & Depth & Speed & $F_r$ \\
(inches) & (inches) & (knots) & \\ \hline
360 & 12.84 & 5 & 0.27163 \\ \hline
360 & 14.33 & 7 & 0.38028 \\ \hline
360 & 15.31 & 8 & 0.43461 \\ \hline
360 & 15.55 & 9 & 0.48894 \\ \hline
\end{tabular}
\end{center}
\caption{\label{table:transom} Details of transom model tests.}
\end{table}

\begin{figure}
\begin{center}
\includegraphics[width=\linewidth]{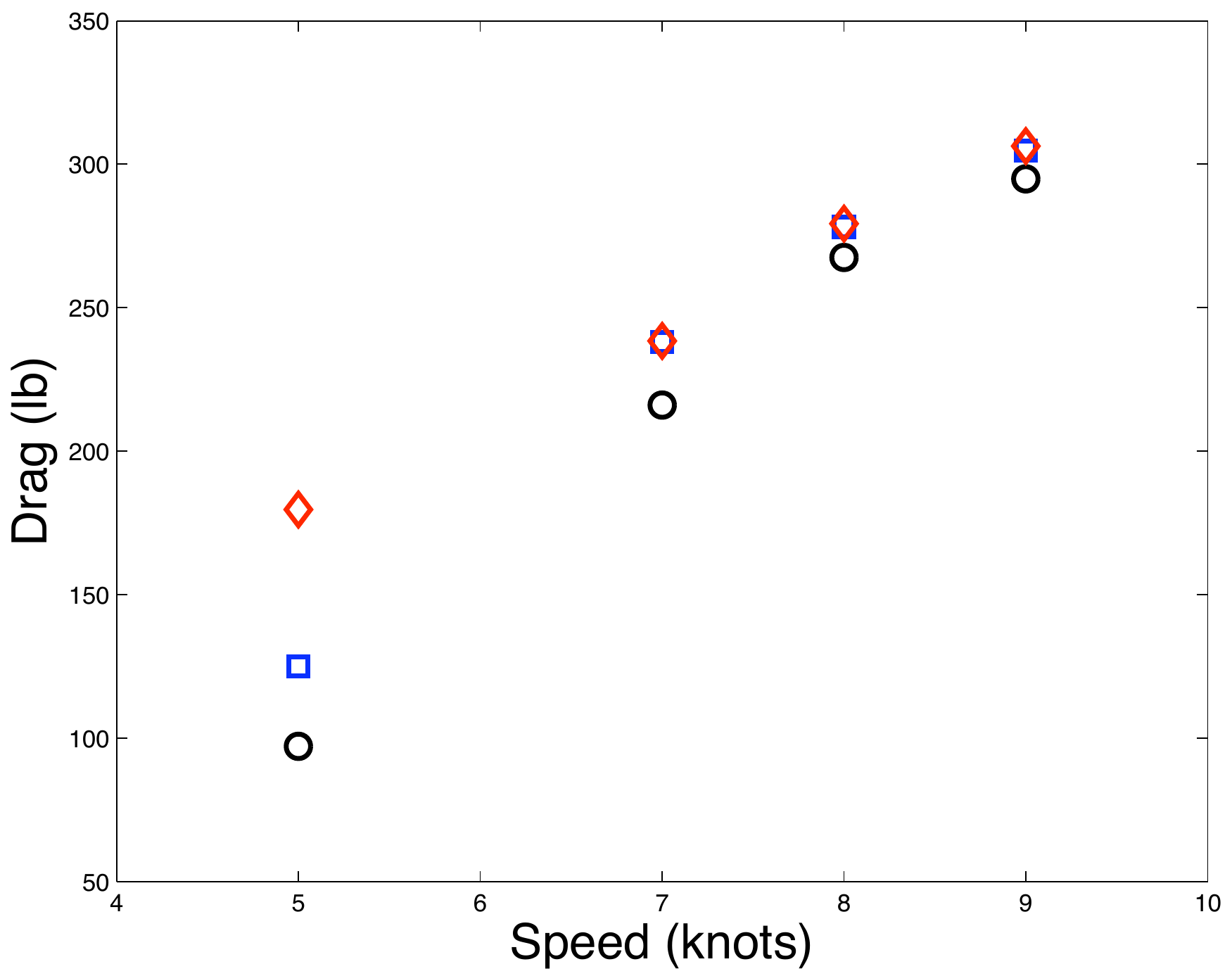}
\end{center}
\caption{\label{drag} Drag of transom-stern model as a function of speed.  DTMB measurements, NFA predictions, and Das Boot predictions are respectively denoted by black circular, blue square, and red diamond symbols.}
\end{figure}

Figure \ref{drag} compares DTMB measurements of the transom-stern model's drag as a function of speed to NFA and Das Boot predictions.   The NFA and Das Boot force predictions are respectively based on formulations that are provided in \citeasnoun{dommermuth07} and \citeasnoun{wyatt00}.  We note that Das Boot uses a Hughes form factor of 0.08.  For speeds greater than five knots, NFA and Das Boot predictions agree with each other and are very close to DTMB measurements.    At five knots, the predicted drags are higher than the measured drag.  This could indicate that the numerical methods have inadequate recovery of the pressure in the stern region.   Research is ongoing in this area. 

\begin{figure*}
\begin{center}
\begin{tabular}{lll}
(a) & & \vspace{-12pt} \\
& \includegraphics[width=0.35\linewidth]{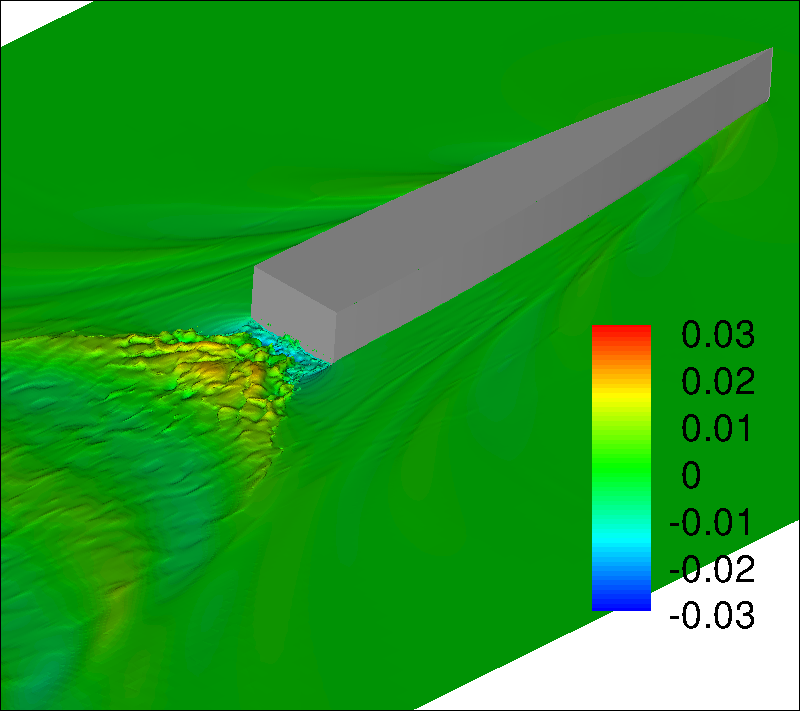}
& \includegraphics[width=0.4\linewidth]{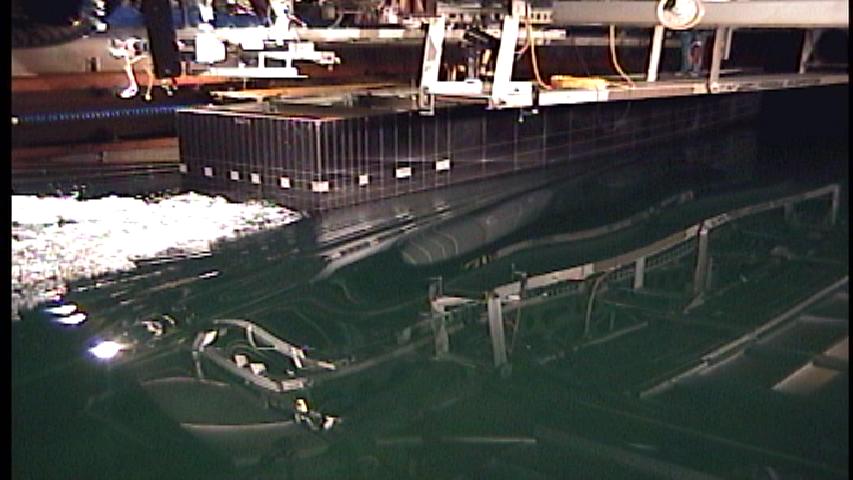} \\
(b) & & \vspace{-12pt} \\
& \includegraphics[width=0.35\linewidth]{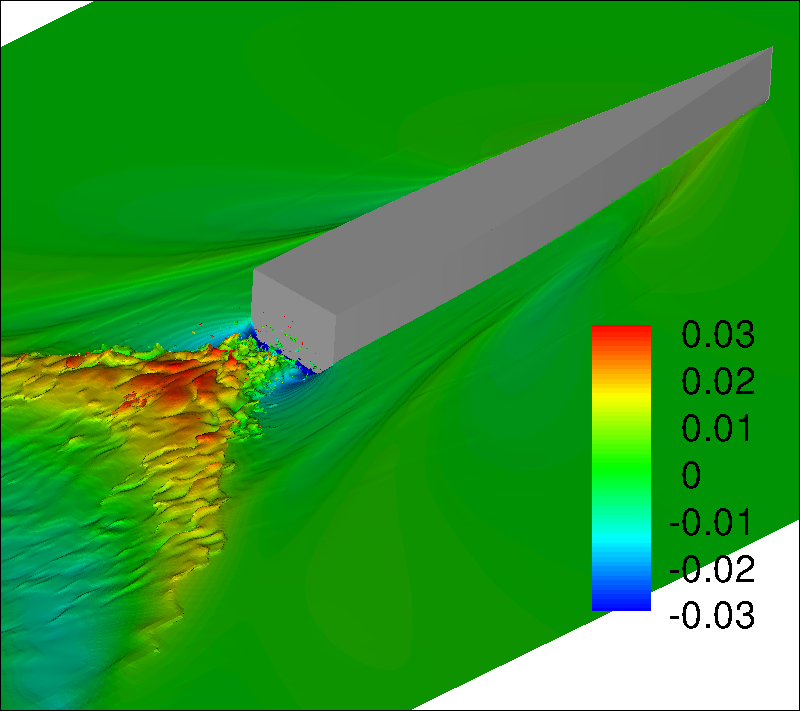}
& \includegraphics[width=0.4\linewidth]{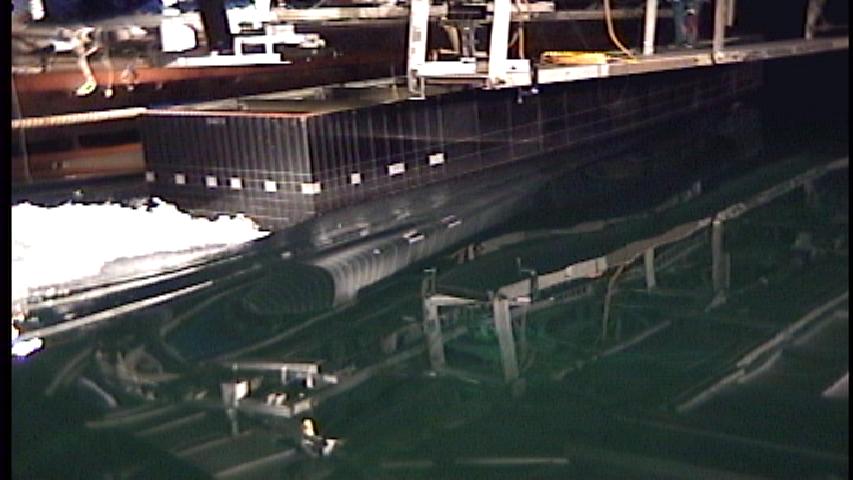} \\
(c) & & \vspace{-12pt} \\
& \includegraphics[width=0.35\linewidth]{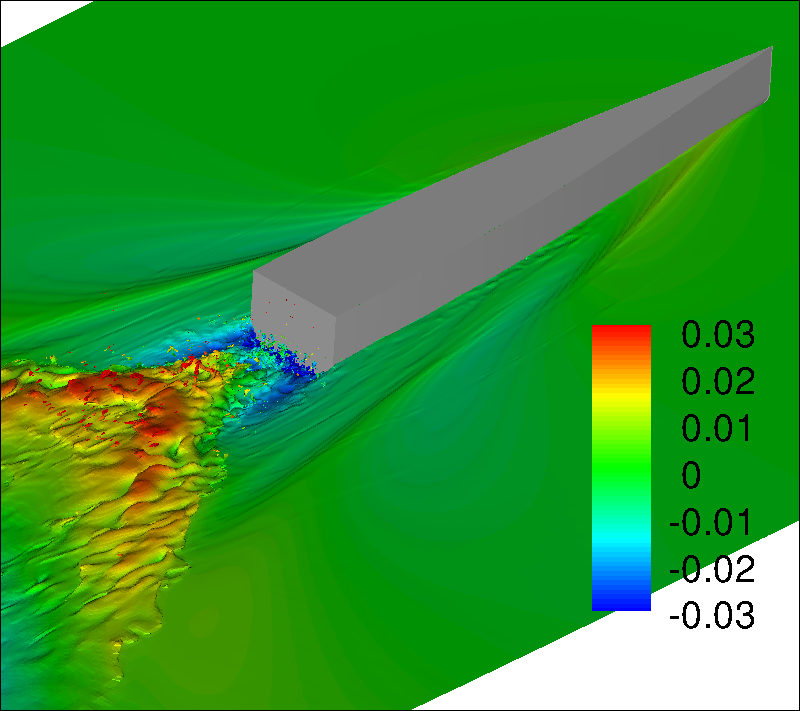}
&\includegraphics[width=0.4\linewidth]{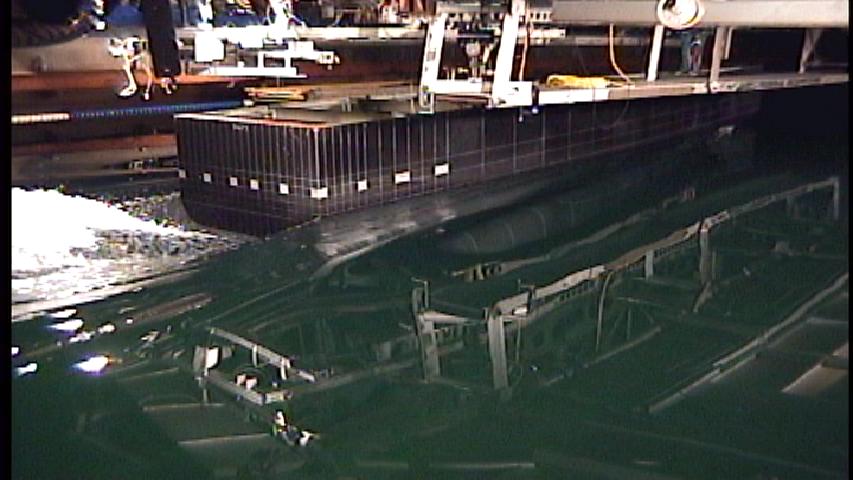} \\
(d) & & \vspace{-12pt} \\
& \includegraphics[width=0.35\linewidth]{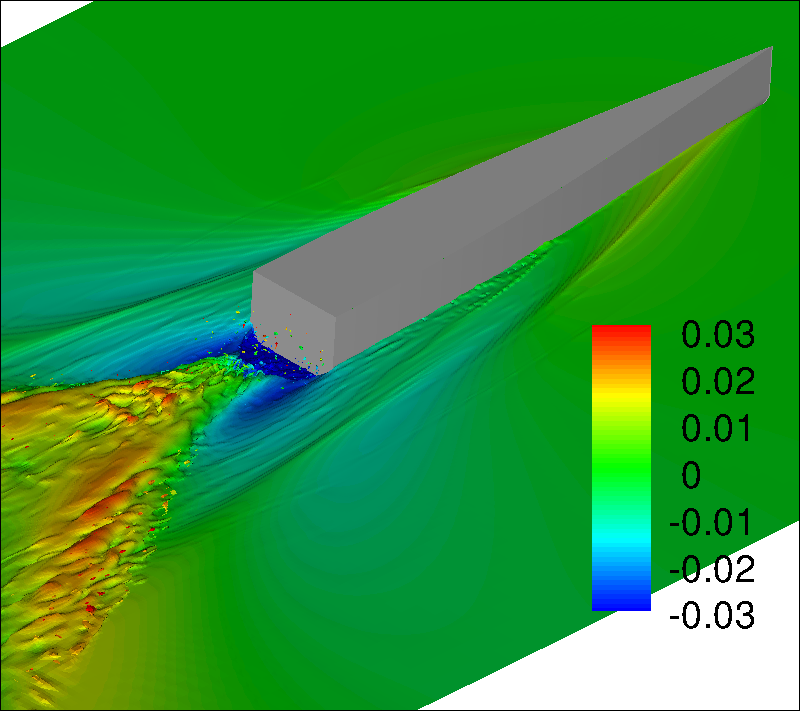}
& \includegraphics[width=0.4\linewidth]{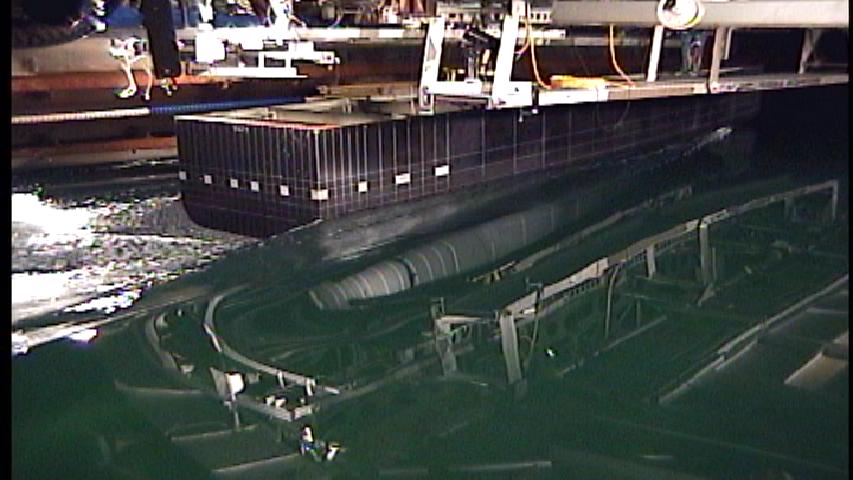}
\end{tabular}
\end{center}
\caption{\label{perspect} Transom-stern perspective views.  (a) 5 knots, (b) 7knots, (c) 8 knots, and (d) 9 knots.  NFA predictions and DTMB measurements are respectively shown on the left and right.}
\end{figure*}

Figure \ref{perspect} (a-d) show perspective views of measured and predicted free-surface elevations in the transom-stern region.  The numerical results are shown at non-dimensional time $t=5$.  The transom is wet when the towing speed is five knots, and dry at nine knots.   Glassy regions immediately behind the transom are evident at speeds of eight and nine knots.   On average, the rooster tails are higher for the two intermediate speeds, seven and eight knots.
  
\begin{figure*}
\begin{center}
\begin{tabular}{llll}
(a) & & (b) & \vspace{-15pt} \\
& \includegraphics[width=0.4\linewidth]{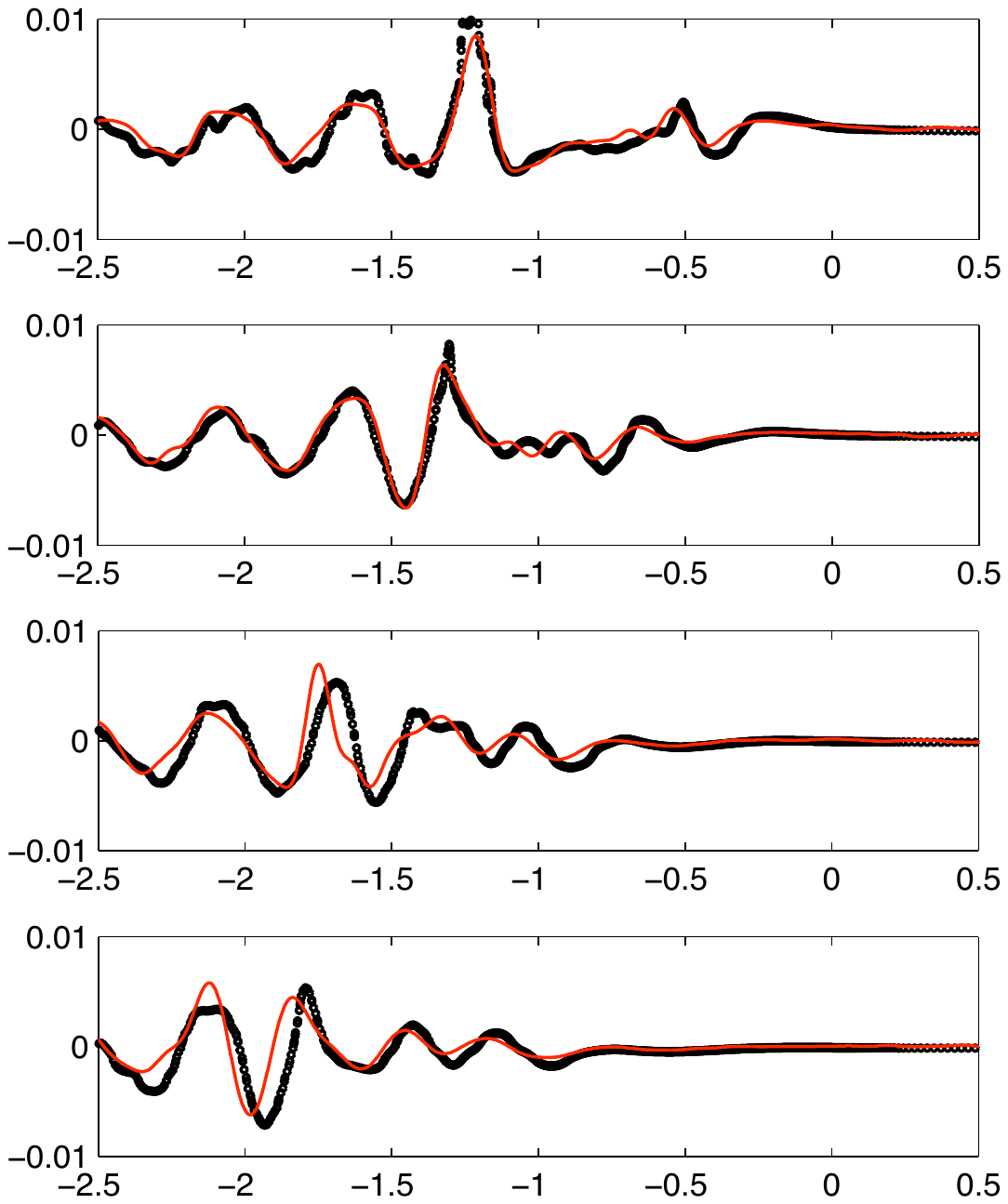}
& & \includegraphics[width=0.4\linewidth]{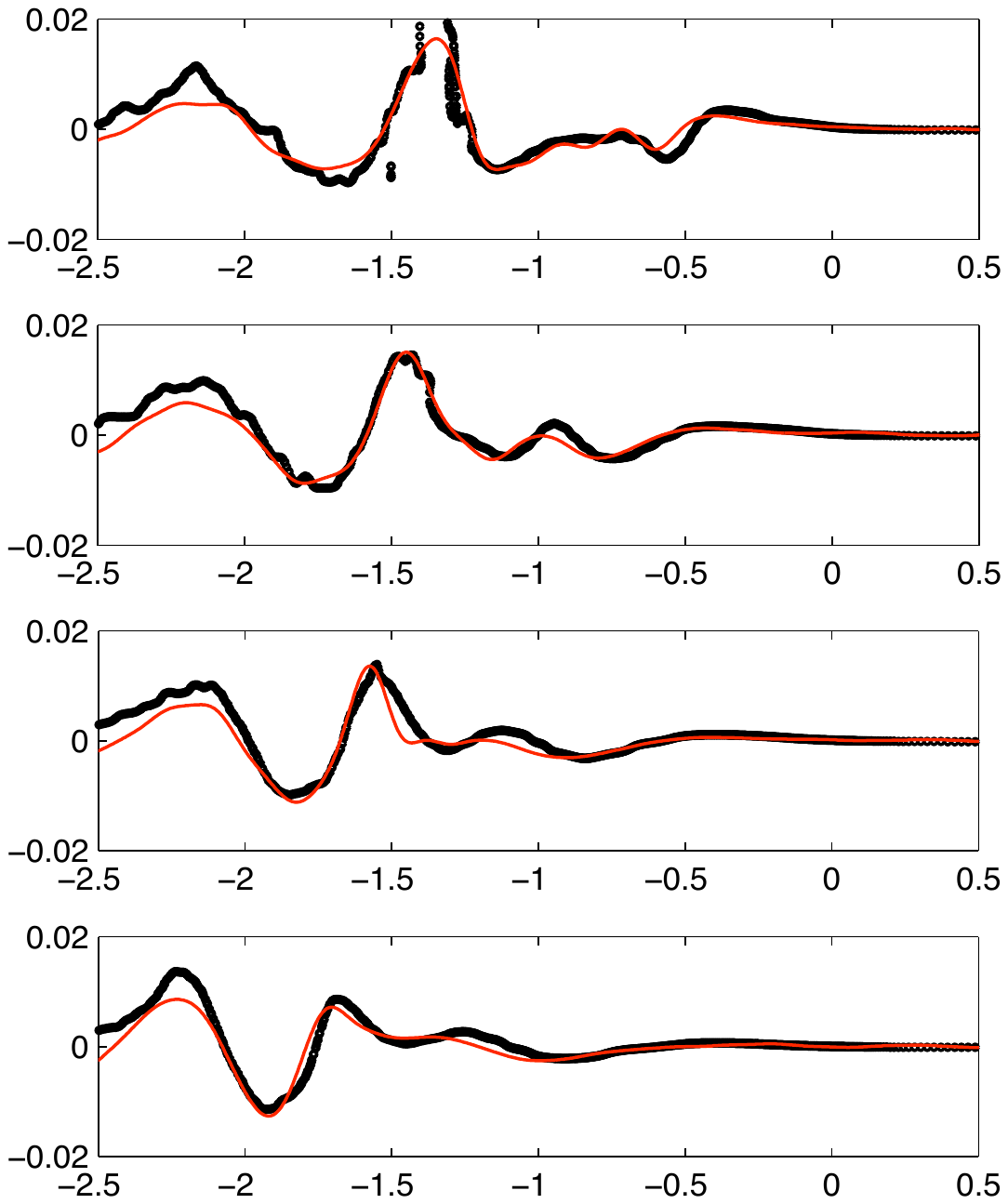} \\
& & & \\
(c) & & (d)  \vspace{-15pt} \\
& \includegraphics[width=0.4\linewidth]{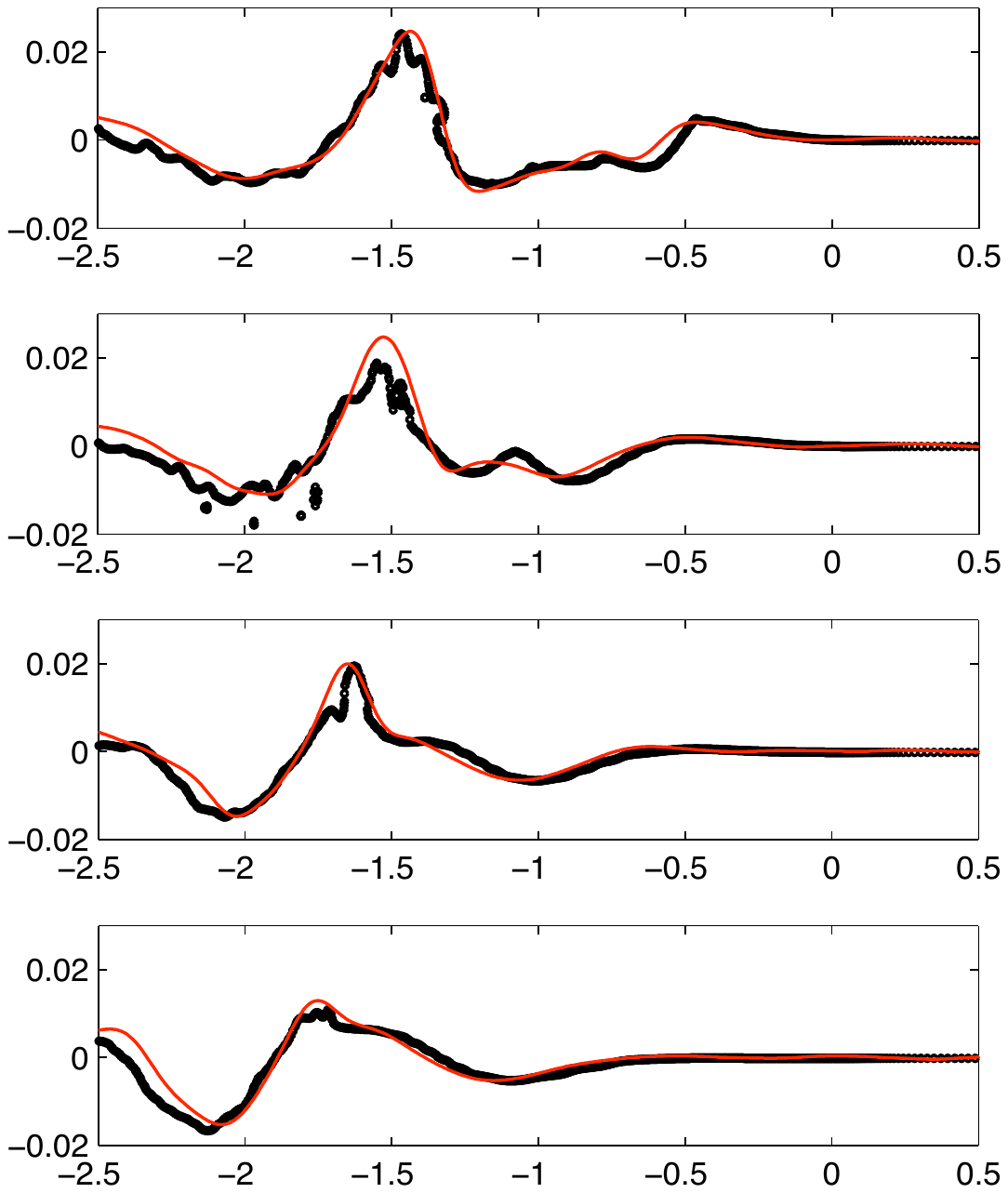}
& & \includegraphics[width=0.4\linewidth]{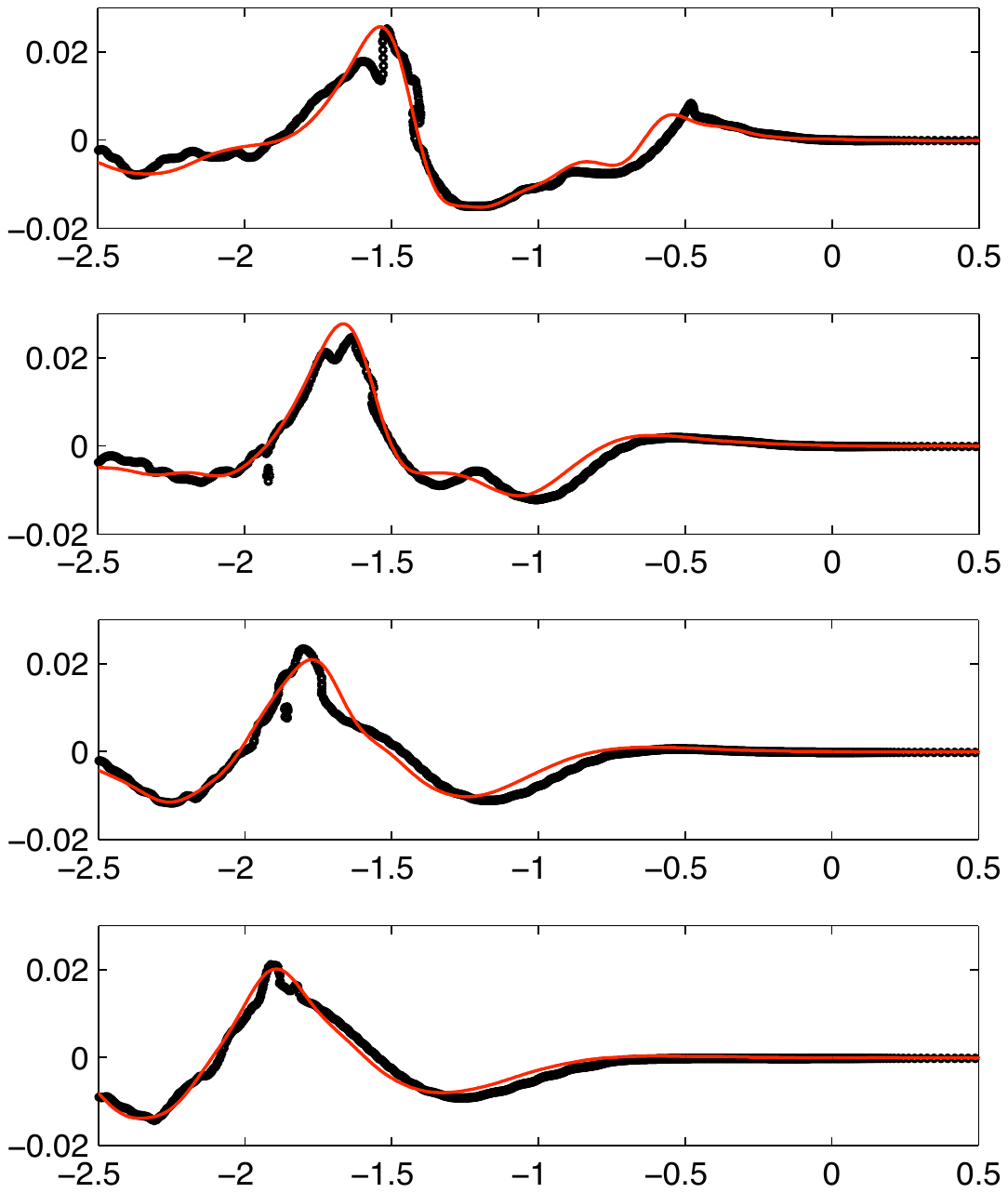}
\end{tabular}
\end{center}
\caption{\label{wavecut} Transom-stern model wave cuts.  (a) 5 knots, (b) 7knots, (c) 8 knots, and (d) 9 knots.   DTMB measurements (red lines) are compared to NFA predictions (black lines).  For each speed and from top to bottom, the transverse cuts are located at $y/L_o$=0.14375, 0.22847, 0.3125, and 0.39514}
\end{figure*}

Figure \ref{wavecut} (a-d) compares DTMB measurements and NFA predictions of wave cuts for four speeds and four transverse locations.  The numerical results are shown at non-dimensional time $t=5$.  The agreement between measurements and predictions is very good for all four speeds and locations.   Some wave breaking and air entrainment is evident in the numerical predictions. 

\section{Conclusions}

In terms of progress, it is interesting to consider the results of research reported in earlier ONR symposiums.   \citeasnoun{dommermuth98} study the flow near the bow of model 5415 using a variable-density, cartesian-grid formulation.   A body force is used by \citeasnoun{dommermuth98} to impose the body boundary condition.   The numerical results of     \citeasnoun{dommermuth98}  barely capture the initial onset of wave overturning near the bow.  \citeasnoun{sussman01} continue to develop interface capturing methods.  Once again, comparisons are shown to the bow flow of model 5415.   The results do not show significant improvement over their earlier results.  However, their calculations of the breakup of a turbulent spray sheet illustrate a novel application of interface-capturing methods.  \citeasnoun{dommermuth04} use two methods to study the flow around model 5415, a vertical strut, and a bluff wedge.  The first method uses free-slip conditions on the hull in combination with a hybrid level-set and VOF interface-capturing method.   In addition,  adaptive mesh refinement (AMR) is used to improve grid resolution near the hull and free-surface interface.   Their preliminary results illustrate the efficiency  of AMR.  The second method uses body-force and VOF formulations on a cartesian grid with no grid stretching.  The results show more fine-scale detail than the earlier studies.  The predicted free-surface elevations compare well with experiments, but the body-force method is too ``sticky" because too much fluid is dragged with the ship hull.  Based on these results, \citeasnoun{dommermuth06} use free-slip boundary conditions to impose the body boundary condition to reduce stickiness.   The VOF algorithm is generalized to include free-slip conditions on the ship hull.  The grid is stretched along the cartesian axes to improve grid resolution.   Numerical predictions compare well with laboratory measurements of ship models moving with constant forward speed.   \citeasnoun{fu08} perform preliminary studies of wave impact.   NFA predictions of a plunging breaking wave agree very well with measurements for free-surface elevations and water-particle velocities.     \citeasnoun{ratcliffe08} continue their earlier studies of incident waves interacting with towed model.    Numerical predictions of the diffracted waves agree with QViz measurements in the bow region for a high speed case with a large incident wave.   \citeasnoun{wyatt08} compare full-scale lidar measurements of the rooster-tail region behind the R/V Athena to numerical predictions using NFA.     Mean free-surface elevations are in good agreement, but free-surface spectrums differ at higher frequencies.    Aliasing may have adversely affected their numerical results.   Also, unlike the full-scale ship, their numerical model is not appended.

The present research shows preliminary comparisons between measurements and predictions for a transom-stern model towed at four different speeds.   The comparisons include drag and wave cuts.    There is good agreement between numerical predictions and laboratory measurements.   Future research will focus on establishing convergence and detailed studies of the transom-stern region, including wave breaking, spray formation, and air entrainment.   

\section{Acknowledgements}

The Office of Naval Research supports this research under contract number N00014-07-C-0184. Dr. Patrick Purtell is the program manager.  This work is supported in part by a grant of computer time from the DOD High Performance Computing Modernization Program (http://www.hpcmo.hpc.mil/).  The numerical simulations have been performed on the Cray XT3 at the U.S. Army Engineering Research and Development Center (ERDC). 

\bibliography{27onr}
\bibliographystyle{27onr}

\end{document}